STUDIES

Paolo Molaro

# ON THE EARTHSHINE DEPICTED
# IN GALILEO'S WATERCOLORS OF THE MOON

SUMMARY

With the manuscript of the *Sidereus Nuncius* preserved at the Biblioteca Nazionale of Florence are included 7 watercolors of the Moon painted by Galileo himself. We suggest that some of them, and in particular the drawing of the 30 Nov 1609 of the very first Moon's observations, illustrate the phenomenon of the Earthshine of the Moon, which was discussed in some detail in the *Sidereus Nuncius* to provide evidence of the similarity of Earth to other celestial bodies. The watercolors were used as models for the engraving of the Moon in the *Sidereus* but, surprisingly, the secondary light had not been reproduced. Galileo may have decided for the inclusion of the passage on the Earthshine only at a very late stage of the editorial process. Galileo's hesitation shows how contentious was this issue already recognized as a possible discriminant between the different systems of the world.

*Keywords*: Galileo Galilei, *Sidereus Nuncius*, Moon, Secondary light.

1. Etchings and watercolors

In the *Sidereus Nuncius* published in mid-March 1610 there are five etchings of the Moon, that Galileo inserted to illustrate what captured by the first telescopic observations. They have been extensively studied with particular attention to the dates of the observations and to the identification of the features which are visible on the illuminated part of the Moon or along the terminator.[1]

INAF-OAT - Via G.B. Tiepolo, 11 - 34143 Trieste, Italy.

[1] See Guglielmo Righini, *New Light on Galileo's Lunar Observations*, in Maria Luisa

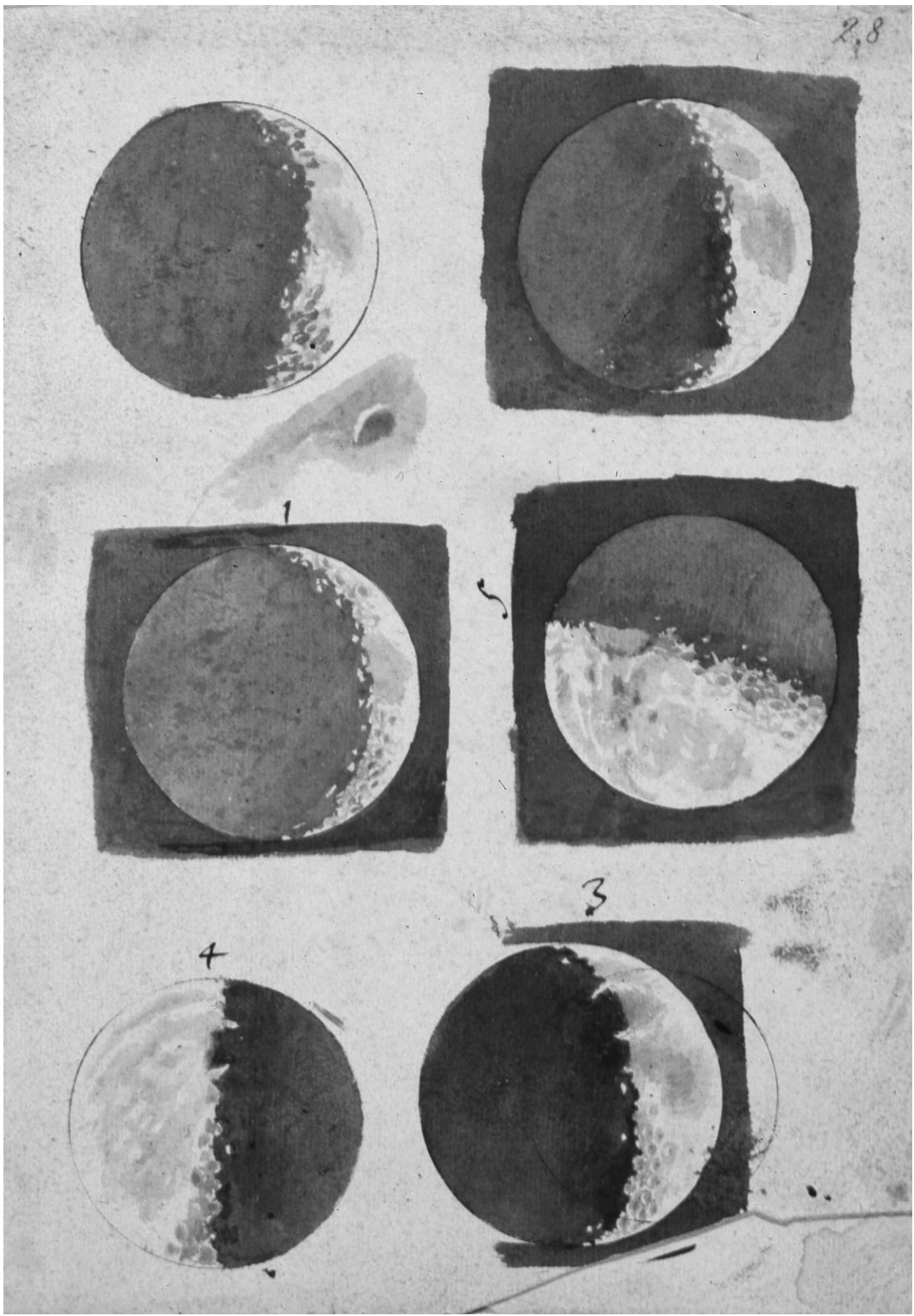

Fig. 1. BNCF, Ms. Gal. 48, f. 28r. Courtesy of Ministero per i Beni e Le attività Culturali della Repubblica Italiana.

Kopal claimed that drawings are only schematic impressions of the lunar surface,[2] but Righini showed that they are a faithful representation of what Galileo saw and that he was able to capture relatively small lunar features. Whitaker took also images of the Moon reproducing the phases of Galileo's drawings showing that the observations were performed from 30 November 1609 through January 1610.

Bound with the handwritten manuscript of the *Sidereus Nuncius* preserved at the Biblioteca Nazionale of Florence there are seven wash drawings of the Moon.

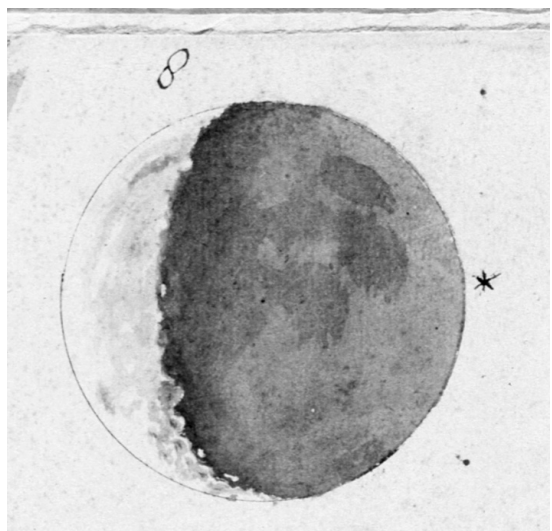

Fig. 2. BNCF, Ms. Gal. 48, f. 29v (detail). Courtesy Ministero per i beni e le attività culturali della Repubblica Italiana.

Six are arranged into two columns on the recto of a page numbered 28, which is shown in Fig. 1, and the seventh is on the verso of a page numbered 29 and shown in Fig. 2. The lower four on bifolio 28 are numbered by Galileo as 1, 5, 4 and 3 and the one on bifolio 29 is numbered as 8. The two sheets were associated with *Sidereus Nuncius* since 1820-1830 by Antinori, and Antonio Favaro accepted this connection but without the horoscope which appears also on the same page.

The horoscope on the verso of page 29 is for May 2 1950, which is the birth date of Cosimo II de Medici, and might reveal the intention of Galileo to dedicate the book to the Medici;[3] a fact that is consistent with the water-

---

RIGHINI BONELLI – WILLIAM R. SHEA (eds.), *Reason, Experiment and Mysticism in the Scientific Revolution*, New York, Science History Publications, 1975, pp. 59-76; OWEN GINGERICH, *Dissertatio cum Professore Righini et Sidereo Nuncio*, ibid., pp. 77-88; STILLMAN DRAKE, *Galileo's First Telescope Observations*, «Journal for the History of Astronomy», VII, 1976, pp. 153-168; EWAN A. WHITAKER, *Galileo's Lunar Observations and the Dating of the Composition of Sidereus Nuncius*, «Journal for the History of Astronomy», IX, 1978, pp. 155-169; OWEN GINGERICH & ALBERT VAN HELDEN, *From Occhiale to Printed Page: The Making of Galileo's Sidereus Nuncius*, «Journal for the History of Astronomy», XXXIV, 2003, pp. 251-267; EWAN A. WHITAKER, *Mapping and Naming the Moon: A History of Lunar Cartography and Nomenclature*, Cambridge, Cambridge University Press, 1999; EWAN A. WHITAKER, *Identificazione e datazione delle osservazioni lunari di Galileo*, in PAOLO GALLUZZI (a cura di), *Galileo. Immagini dell'universo dall'antichità al telescopio*, Firenze, Giunti, 2009, pp. 263-267.

[2] Cf. ZDENEK KOPAL, *The Earliest Maps of the Moon*, «The Moon», 1969, pp. 59-66.

[3] See GINGERICH, *Dissertatio cum Professore Righini et Sidereo Nuncio* (cit. note 1); GU-

colors painted by Galileo during the preparation of the *Sidereus*. The drawing on the verso of page 29 (Fig. 2) shows the Moon eclipsing a bright star which Whitaker identified with the star $\theta$ *Librae* on 19 January 1610.[4]

Since in that epoch only Galileo could have represented the Moon that way, this has been taken as evidence that the watercolors were painted by Galileo with his own hands.

The watercolors are so carefully painted that it was initially thought that Galileo collected sketches during his observations but drew them subsequently. Gingerich and Van Helden argued instead that the drawings have been made while observing and even proposed the sequence of drawings which follow the temporal observations.[5] Four drawings show the Moon as it appeared in Padova between 30 Nov. and $2^{nd}$ of Dec. 1609. Other two on December 17-18 and the seventh, the one with $\theta$ *Librae*, on 19 Jan. 1610.

The Moons are drawn as circles of about 57-59 mm in diameter and made with a compass whose center can be even seen as a tiny dot. The base color of the paper is used for the lighted area of the Moon illuminated by the Sun while sepia ink with different densities is used to obtain a beige variations to illustrate the physical structure of the Moon.

Galileo uses the light-dark contrast to describe the irregular surface of the Moon made of mountains and valleys and in particular to highlight the white spots in the dark regions close to the terminator.

The watercolors were probably used someway for the etchings reproduced in the *Sidereus Nuncius*, although there is not a one-to-one correspondence.[6] Gingerich and Van Helden argued that Galileo produced the second etching, the one of the first quarter Moon, from the watercolor n. 3 copying the pattern of light and dark seen along the terminator at rather different phases. The mistery of the *missing 2* has been explained by Gingerich who provided evidence that Galileo pressed for space combined two images into one thus reinforcing the idea that the watercolors are indeed the Galileo's source for the reworked etchings.[7]

Yet there are significant differences between the etchings and the watercolors: the most striking of which is certainly that none of them reproduce

---

GLIELMO RIGHINI, *L'oroscopo galileiano di Cosimo II de' Medici*, «Annali dell'Istituto e Museo di Storia della scienza di Firenze», I, 1976, pp. 28-36.

[4] See WHITAKER, *Galileo's lunar observations* (cit. note 1).

[5] See GINGERICH & VAN HELDEN, *From Occhiale to Printed Page* (cit. note 1).

[6] For a different view see HORST BREDEKAMP, *I molteplici volti del Sidereus Nuncius*, in LUCIA TONGIORGI TOMASI – ALESSANDRO TOSI (a cura di), *Il cannocchiale e il pennello: nuova scienza e nuova arte nell'età di Galileo*, Firenze, Giunti, 2009, pp. 169-174.

[7] See OWEN GINGERICH, *The Mystery of the Missing 2*, «Galilæana», IX, 2012, pp. 91-101.

the perfectly round crater as it appears on the second and third Moons in *Sidereus*. According to Gingerich the feature is essentially an archetypal view of the illumination of a crater added to the Moon image to save other two planned distinct figures. A second important one and, so far generally unnoticed, is in the background; this remarkable difference is the subject of this discussion and we argue below that it carries specific meaning.

2. SECUNDA LUNAE CLARITAS

The first Moon watercolor on the page is drawn against the base color of the page. In this representation the drawing focuses on the internal structure of the irregular line of the terminator and of the structure on the illuminated sector. The successive three drawings are reproduced onto a dark background. This kind of representation emphasizes the illuminated part of the crescent Moon. This is the case also in the sixth image (n. 3) where the dark background is depicted only in correspondence of the bright part of the Moon. A dark background depicted all around the Moon is unnecessary for the illustration, but it becomes a mean to emphasize the different degrees of darkness of the non-illuminated part of the Moon against the dark sky, namely the phenomenon of the Earthshine (called also secondary or ashen light). Galileo describes the Secunda Lunae Claritas or Earthshine in a passage of the *Sidereus*:

a certain faint light is also seen to mark out the periphery of the dark part which faces away from the Sun, separating this from the darker background of the aether.[8]

The Moon watercolors with dark background depict the Moon's night side as being bathed in a soft, faint light. This is particularly evident in the drawing N. 1 of the crescent Moon.

There is a passage in the *Sidereus Nuncius* which explains why the Earthshine is stronger for the crescent Moon, is at its minimum in the decreasing Moon and is faint at quadrature. The passage seems inspired by the watercolors:

Moreover, it is found that this secondary light of the Moon is greater according as the Moon is closer to the Sun. It diminishes more and more as the Moon recedes from that body until, after the first quarter and before the last, it is seen very weakly and uncertainly even when observed in the darkest sky. But when the Moon is within sixty degrees of the Sun it shines remarkably, even in twilight; so brightly indeed that with the aid of a good telescope one may distinguish the large spots.[9]

---

[8] OG, III, p. 72. English trans. in STILLMAN DRAKE (ed.), *Discoveries and Opinions of Galileo*, New York, Anchor Books, 1957, p. 42.

The last sentence is particularly adapted to interpret the *maria* depicted on the dark side of the Moon as darker shadows in the drawing n. 8 on the verso of folio 29 the one with the $\theta$ *Librae*. The *maria* are just sketched to illustrate the effect and probably added later since they do not reproduce the angular position between them and the terminator correctly. The need to highlight the star could explain why this watercolor is the only one with the Moon in the sextile and without a dark background. We note that the Moon n. 5 is drawn in a phase where unlikely Earthshine could be detected by naked eye. However, according to above text, it is seen also at quadrature, though very weak.

As was accounted by Panofsky and Bredekamp[10] the young Galileo received comprehensive training as a draftsman, and had prominent Renaissance artists and architects among his best friends acquiring an aesthetic mentality. In 1588 he attained an instructor position in the Accademia delle Arti del Disegno in Florence, teaching perspective and chiaroscuro and Salusbury (1664) in his first biography of Galileo made also reference to the existence of a self-portrait.[11] Late in life Galileo told Viviani that if he could have pursued any profession, he would have been a painter. When Galileo wrote to Belisario Vinta in the attempt to enter the service of Cosimo II, the Grand Duke of Tuscany, as court mathematician and philosopher, he mentioned that he had a treatise *De Visu et Coloribus*, but unfortunately no such work have been preserved. Therefore, Galileo's Moon draws are very clear as to the features they were meant to exemplify and nothing is left to chance. The images drawn against a dark background the non-illuminated part looks slightly brighter than the sky and this cannot be by mere chance.

---

[9] OG, III, p. 73. English trans. in DRAKE (ed.), *Discoveries and Opinions of Galileo* (cit. note 8), p. 42.

[10] ERWIN PANOFSKY, *Galileo as a Critic of the Arts. Aesthetic Attitude and Scientific Thought*, «Isis», XLVII, 1956, pp. 3-15; HORST BREDEKAMP, *Galilei der Kunstler: Der Mond, die Sonne, die Hand*, Berlin, Akademie Verlag, 2007.

[11] See JOHN ELLIOT DRINKWATER, *Life of Galileo Galilei with Illustrations of the Advancements of Experimental Philosophy*, London, Clowes, 1829, p. 103; see also NICK WILDING, *The return of Thomas Salusbury's Life of Galileo (1664)*, «British Journal for the History of Science», XLI, 2008, 2, pp. 241-265.

The text in the *Sidereus* appears closer to watercolors than to the etchings where there is nothing to illustrate Galileo's words on the Earthshine. It is likely that when Galileo was drafting the description of his observations of the Moon, he had not yet fully settled in his mind how far to go with the issue of the Earthshine and what details to present in the book's illustrations.

3. ON THE INTERPRETATION OF THE EARTHSHINE

The phenomenon of the Earthshine that the crescent Moon displays on its dark side is addressed by Galileo in the *Sidereus Nuncius* following the description of the physical nature of the Moon:

I like this point to cite the cause of another lunar phenomenon worthy of admiration, that although I have not seen recently but many years ago and showed, explained and clarified in his cause to some intimate friends and disciples.[12]

He shows that the Earthshine cannot be produced by light from Venus or by a semitransparent Moon but that it is produced by the sunlight reflected by the Earth towards the crescent Moon. Galileo defers dealing with it in more detail to the treatise *De Systemate Mundi*:

But these few remarks suffice on the subject: this will be discussed more fully in our *De Systemate Mundi*, where with many very valid reason and experience will show the reflection of Sunlight from the Earth to those who are saying you should exclude it from the list of errant stars especially since it has no motion and light, and we shall show that it revolves around and has a greater splendor than the Moon.[13]

Nor was Galileo the first to analyze the phenomenon. The first ones to write about it were Leonardo da Vinci and Paolo Sarpi, probably independently from each other. Leonardo da Vinci wrote about it already in the period 1506-1509 in his notebook I:

---

[12] Here I wish to assign the cause of another lunar phenomenon well worthy of notice. I observed this not just recently, but many years ago, and pointed it out to some of my friends and pupils, explaining it to them and giving its true cause.

[13] Let these few remarks suffice us here concerning this matter, which will be more fully treated in our System of the world [original: De Systemate Mundi]. In that book, by a multitude of arguments and experiences, the solar reflection from the earth will be shown to be quite real-against those who argue that the earth must be excluded from the dancing whirl of stars for the specific reason that it is devoid of motion and of light. We shall prove the earth to be a wandering body surpassing the moon in splendor, and not the sink of all dull refuse of the universe; this we shall support by an infinitude of arguments drawn from nature.

some have believed that the Moon has some light of its own, but this opinion is false, for they have based it upon that glimmer visible in the middle between the horns of the new Moon...this brightness at such a time being derived from our ocean and the other inland seas, for they are at that time illuminated by the Sun, which is then on the point of setting, in such a way that the sea then performs the same office for the dark side of the Moon as the Moon when at full does for us when the Sun is set.[14]

However, Leonardo's notebooks were kept by Francesco Melzi and the circulation was delayed until 1585-1590 and published in 1597. The manuscripts almost certainly circulated in the circle of Pinelli as well as among Florentine painters and could have influenced Galileo's first reflections on the light of the Moon.

In the years 1578-1583, before Leonardo's manuscripts were circulated Fra' Paolo Sarpi wrote about the Earthshine in "pensiero n. 28" of his *Scritti filosofici e Teologici*:

Because of the seas, which have a polished surface, the Earth gives more light to the Moon than does the Moon to the Earth, and that light that we see on the darkened part of the Moon's face when she is crescent comes, perhaps, from the Earth, since it cannot be the Moon's own.[15]

Galileo met Sarpi shortly after his arrival in Padua in 1592 and it is quite plausible that they discussed the issue of the Earthshine. Sarpi's contribution on this matter is not acknowledged by Galileo in the Sidereus Nuncius and the twist in their relationship after 1610 is well documented in the books by Reeves and by Bucciantini, Camerota, Giudice.[16]

Kepler in his *Astronomia pars optica* of 1604 cited an interpretation of the Earthshine published by Michael Maestlin in 1596, in which solar rays reflected from the Earth served to light up the lunar surface. Francesco Patrizi in his *Nova de Universis Philosophia* (1591), a book banned in the following year by the Church, offered also a description, not even totally correct, of the Earthshine that the Moon has at the beginning of the cycle but without identifying the Earth as the origin of the phenomenon. Raffaello

---

[14] LEONARDO DA VINCI, *The Notebooks of Leonardo da Vinci*, arranged, rendered into English and introduced by Edward McCardy, New York, G. Braziller, 1938, I, p. 295.

[15] PAOLO SARPI, *Scritti filosofici e teologici inediti*, a cura di Romano Amerio, Bari, Laterza, 1951, p. 10. English translation in EILEEN REEVES, *Painting the Heavens. Art and Science in the Age of Galileo*, Princeton, Princeton University Press, 1997, p. 33.

[16] See REEVES, *Painting the Heavens* (cit. note 15); MASSIMO BUCCIANTINI – MICHELE CAMEROTA – FRANCO GIUDICE, *Il telescopio di Galileo. Una storia europea*, Torino, Einaudi, 2012, pp. 24-45.

Gualterotti in his *Scherzi degli Spiriti Animali* (1605) provided a correct explanation, which in the same year was opposed by Lodovico delle Colombe who defended the Aristotelian view of the Cosmos.

Galileo returned again on the issue of the Earthshine quite extensively in the *Dialogo* (1632), and again at the end of his life when, already blind, he exchanged opinions with Liceti, who was arguing that the Moon Earthshine was similar to the luminosity of the barite stone and therefore of the same origin. The letter of Galileo was included in the *De lunae suboscura luce prope coniunctiones et in eclipibus observata libri tres* that Liceti published in 1642, shortly after Galileo's death.

We do not know when Galileo actually started to interpret the Earthshine as due to Earth. However, he certainly reached this explanation well in advance to his telescopic observations.

Very likely among Galileo's friends with whom, according to the specific passage in the *Sidereus Nuncius*, he has shown and explained the Earthshine of the Moon is Lodovico Cigoli, that Galileo considered the greatest Tuscan artist of his period. Cigoli was the first painter to depict a physical Moon in his fresco of the *Immacolata* in the church of Santa Maria Maggiore in Rome of 1612. Galileo and Cigoli were close friends since the epoch they both studied perspective in Florence with Ostilio Ricci. There are not many records of that period, but from the letters they exchanged later we may surmise that they maintained a close and friendly relationship for their whole life. As pointed out by Eileen Reeves,[17] the painting of Cigoli could be used to infer the evolution of the concept of the Earthshine in Galileo. Cigoli inserted a crescent Moon in his *Adoration of the Shepherds* made in 1599 (now in Metropolitan Museum of New York), in the *Adoration of the Shepherds* in the church of San Francesco in Pisa (1602), and in the *Deposition* of 1607 (Palatina Gallery, Florence). In this last painting Cigoli depicted a more precise concept of the Earthshine, thus showing that the concept of the Earthshine was taken by Galileo much before the publication of the *Sidereus Nuncius* and likely before Gualterotti. We note that Domenico Tintoretto, who made a famous portrait of Galileo around 1604, also depicted a very similar crescent moon in the background of his Magdalena Penitente (1598-1602, Musei Capitolini).

---

[17] See REEVES, *Painting the Heavens* (cit. note 15).

4. On the absence of the Earthshine in the Moon's Etchings

Nowadays the statement on the Earthshine is not perceived as one of the important items discussed by Galileo in the *Sidereus Nuncius*. However, at the beginning of the XVII century the perception was quite different. For instance, on March 13, 1610, the day of publication of the Galileo's *Sidereus Nuncius*, sir Henry Wotton, the British ambassador in Venice, wrote to his home office about the «*strangest piece of news that hath ever yet received from any part of the world*». Wotton describes Galileo's use of the telescope and his discoveries of the four new planets, the true cause of the Via Lactea, and lastly that «*the Moon is not spherical, but endued with many prominences, and illuminated with the solar light by reflection from the body of the Earth*».[18] According to the Aristotelian philosophy the light reflected by the Earth could not go further than the sub-lunar world.

In turning the telescope to the Moon Galileo certainly was seeking for confirmation of his beliefs. Already in his *Considerations of Alimberto Mauri* (1604), Galileo described the rough surface of the Moon and argued that it does not shine with light of its own but only reflects solar light.

It is interesting to note that the Earthshine is pictured at best in the drawing which was identified as the first done by Galileo already on 30 Nov 1609,[19] or on 1st Dec. 1609 according to Drake.[20] If our interpretation is correct, this shows that Galileo was very interested in the aspect of the Earthshine since his very first Moon observations. The observation of the Earthshine could have been one of the main motivations which lead Galileo to turn its newly developed instrument towards the crescent Moon at the end of November 1609. We emphasize that showing that the Earth was much alike the other celestial bodies was much more revolutionary than showing the other celestial bodies being like the Earth. Thus the Earthshine of the Moon had a specific place against the Aristotelian view of the Cosmos and as such had been described in some detail in the *Sidereus*. As discussed by Reeves[21] and shown by Henry Wotton's letter the concept of Earthshine was seen in much more radical fashion than we would think today. Those who accepted it would have been seen as supporters of a Copernican world system.

---

[18] Henry Wotton to Robert Cecil, March 13, 1610; see L.P. Smith, *The life and letters of Sir Henry Wotton*, Oxford, Clarendon Press, 1907, pp. 485-487: 486-487.

[19] See Whitaker, *Galileo's lunar observations* (cit. note 1); Gingerich & Van Helden, *From Occhiale to Printed Page* (cit. note 1).

[20] Drake, *Galileo's First Telescope Observations* (cit. note 1).

[21] Reeves, *Painting the Heavens* (cit. note 15).

Thus there must be an important reason of why the etchings do not illustrate the Earthshine of the Moon. Galileo's choice of the etching, preferred to the more common Venetian woodcut, and the transfer technique have been analyzed in detail,[22] and technical motivations do not seem plausible. Etchings have a much more limited range of nuances in comparison to the wash drawings but certainly even an inexperienced etcher, would have been able to reproduce the effect of the Earthshine by making thicker signs or by different exposures of the copper plate into acid bath. All the more so, if the etcher were Galileo himself or participated someway in the process, as has been suggested. Examples of the etching potentiality in reproducing different tonalities could be found in the Lorenzo Sirigatti's book *La pratica della prospettiva* (1596)[23] or in the replica of the Elsheimer's paintings by Goudt Hendrick.

The *Sidereus* is a sort of instant book printed in haste without the attention to fine execution of the details in a lap of few weeks as it is clear from the letter by Galileo to Belisario Vinta of the 19 March 1610.

Needham and Gingerich and Van Helden described in detail the making of the *Sidereus Nuncius*.[24] When contacting the printer Tommaso Baglioni at the end of January, Galileo left the first batch of manuscript text including the first part of lunar observations and likely also the drawings for the etchings. Only after 16 February Galileo produced the first batch of Jovian material and four numbered folio leaves (i.e. one signature) were left over for the remaining part on the lunar observations. The material for this initially skipped signature D was made after Galileo received the Vinta's letters of mid February with the Cosimo de Medici's patronage. It includes the explanation of the smooth circular shape of the Moon by an atmosphere, the calculation of the height of the lunar mountains, the Earthshine and the illustrations of the observations of the Pleiades and Orion, made on 31 January and on 7 February respectively, showing the posi-tion of the new stars. The last things added to the text were the Jupiter observations till the $2^{nd}$ of March, the closure to the book and the dedication to Cosimo de Medici, which was written on the 12 March. Wootton noted that preface, closure and the passage with the description of the earthshine are the only passages in the book in which Galileo declares his support for Copernicanism and, in particular, the earthshine is the only one against the Tycho Brahe's model of the solar system.[25]

---

[22] See IRENE BRÜCKLE and RUTH TESMAR, *Modeling the Transfer*, in IRENE BRÜCKLE and OLIVER HAHN (eds.), *Galileo's Sidereus Nuncius: A Comparison of the Proof Copy (New York) with Other Paradigmatic Copies*, Vol. 1 of HORST BREDEKAMP (ed.), *Galileo's O*, Berlin, Akademie Verlag GmbH, 2011, pp. 98-104; HANS JACOB MEIER, *Galileo's break with the Venetian woodcut*, ibid., pp. 110-114.

[23] See ROBERT FELFE, *Performance of the etchings*, in BRÜCKLE and HAHN (eds.), *Galileo's Sidereus Nuncius* (cit. note 22), pp. 115-124.

[24] See PAUL NEEDHAM, *Galileo Makes a Book: The First Edition of Sidereus Nuncius, Venice 1610*, Vol. 2 of BREDEKAMP (ed.), *Galileo's O* (cit. note 22); GINGERICH & VAN HELDEN, *From Occhiale to Printed Page* (cit. note 1).

The last Moon etching in the Sidereus is a duplicate of the second one. Gingerich provided evidence that this was deliberate,[26] but we note that the place of the duplicate Moon would have been also an appropriate place to illustrate the phenomenon of the earthshine, which was later postponed by Galileo at the end of the Moon observations and before the description of the discovery of new stars.

Thus, it is possible that Galileo initially was not completely confident to illustrate the theory of the origin of the Earthshine effect and then, when he changed his mind and added the relative passage, it was too late to add new etchings in the signature D or modify the previous ones. Galileo's hesitation and extreme caution in dealing with the Earthshine testifies how controversial and crucial was this issue at the beginning of the seventeenth century, already recognized by him as a real physical evidence able to discriminate between the different systems of the world and in particular against a Ptolemaic interpretation of the Tychonian model. Something that will be taken up again and discussed quite extensively in the Galileo's *Dialogo* of 1632.

---

[25] DAVID WOOTTON, *Galileo Watcher of the Skies*, New Haven, Yale University Press, 2010, pp. 103-104.

[26] GINGERICH, *The Mystery of the Missing 2* (cit. note 7).